\documentclass[aps,reprint,12pt]{revtex4-1}
\usepackage[version=3]{mhchem} 
\usepackage[hidelinks]{hyperref}
\usepackage{graphicx}
\usepackage{amsfonts,amsmath,amssymb,bm}
\usepackage{color}
\usepackage{multirow}
\usepackage[utf8]{inputenc}
\usepackage{wrapfig}
\usepackage{comment}
\usepackage{xfrac}
\usepackage{bm}
\graphicspath{{./figures/}}
\usepackage{ragged2e}

\begin{document}
\title{Effect of Strain on the Band Gap of Monolayer MoS$_2$}

\author{Raj K. Sah}
\affiliation{Department of Physics, Temple University, Philadelphia, PA 19122}

\author{Hong Tang, Chandra Shahi, Adrienn Ruzsinszky, and John P. Perdew}
\affiliation{Department of Physics and Engineering Physics, Tulane University, New Orleans, LA 70118}

\begin{abstract}
Monolayer molybdenum disulfide ($\mathrm{MoS_2}$) under strain has many interesting properties and possible applications in technology. A recent experimental study examined the effect of strain on the bandgap of monolayer $\mathrm{MoS_2}$ on a mildly curved graphite surface, reporting that under biaxial strain with a Poisson's ratio of 0.44, the bandgap decreases at a rate of 400 meV/\% strain. In this work, we performed density functional theory (DFT) calculations for a free-standing $\mathrm{MoS_2}$ monolayer, using the generalized gradient approximation (GGA) PBE, the hybrid functional HSE06, and many-body perturbation theory with the GW approximation using PBE wavefunctions (G0W0@PBE). For the unstrained monolayer, we found a standard level of agreement  for the bandgap between theory and experiment. For biaxial strain at the experimental Poisson's ratio, we found that the bandgap decreases at rates of 63 meV/\% strain (PBE), 73 meV/\% strain (HSE06), and 43 meV/\% strain (G0W0@PBE), which are significantly smaller than the experimental rate. We also found that PBE predicts a similarly smaller rate (90 meV/\% strain) for a different Poisson's ratio of 0.25. Spin-orbit correction (SOC) has little effect on the gap or its strain dependence. The strong disagreement between theory and experiment may reflect an unexpectedly strong effect of the substrate on the strain dependence of the gap. Additionally, we observed a transition from a direct to an indirect bandgap under strain, and (under an equal biaxial strain of 10\%) a semiconductor-to-metal transition, consistent with previous theoretical work.
\end{abstract}
\maketitle
\noindent
\section{Introduction}
Graphene is a typical two-dimensional (2D) layered material. Monolayer graphene has been found to exhibit several notable features, including high electrical conductivity, high transparency, high thermal conductivity at ambient temperature, a high Young's modulus, and a high specific surface area \cite{graphene-allen,graphene-Soldano}. Graphene possesses outstanding electrical properties; however, being a zero-gap material \cite{graph-zerogap} limits its application in logical circuits. In an effort to overcome the limitations of graphene and increase its variety of uses, scientists have turned their attention back to alternative 2D materials that resemble graphene \cite{Fewlayer-MoS2,2D-MoS2}. Monolayer MoS$_2$ has gained significant interest over the past decade.

Monlolayer MoS$_2$ has a 2D nature and a large band gap. Similar to graphene \cite{Lee-2008}, a single-layer MoS$_2$ has strong inherent tensile strength and flexibility out of plane, allowing the film to withstand strains of up to $11\%$ before rupturing \cite{Bartolazzi-2011}. According to theoretical predictions, the band gap of semiconducting transition-metal dichalcogenides (TMDCs) would be significantly impacted by tensile strain, and applying approximately 10\% biaxial strain might potentially close the gap entirely \cite{Johari-2012,Yue-2012,Shi-2013}. Strains can be used to control the electronic properties of two-dimensional (2D) materials, which is important for implementation of a 2D material into flexible electronics and next-generation strain engineering devices. Therefore, it becomes important that the material withstands the desired strain, and that we know the effect of the strain on the material. A recent experimental study
of an $\mathrm{MoS_2}$ monolayer on a macroscopically curved surface (with a radius of curvature greater than 10$^7$ Å) has measured a relationship between band gap change and strain in monolayer MoS$_2$ \cite{DJ_Trainer_MoS2}. Trainer \textit{et al.} \cite{DJ_Trainer_MoS2} determined the Poisson's ratio of monolayer MoS$_2$ to be 0.44. A positive Poisson's ratio has been previously reported \cite{Kang2015,Woo-2016}. Because of a positive Poisson's ratio, the crystal will respond to a tensile strain in one direction by compressing in the perpendicular direction. Trainer \textit{et al.} \cite{DJ_Trainer_MoS2} applied a tensile strain along an in-plane direction and found that the quasiparticle band gap decreases at the rate of 400 meV/\% strain until a nominal strain of 2.5\% and at a much slower rate from 2.5\% to 4.9\%. In this work, we have studied band gap change with strain for a free-standing $\mathrm{MoS_2}$ monolayer theoretically, using the Heyd-Scuseria-Ernzerhof screened hybrid functional HSE06 \cite{HSE06-Krukau-2006} and G0W0 \cite{GW-Hedin-1965,GW-Hypersten-1985,GW-schif-2006} methods. We also used the Perdew-Burke-Ernzerhof (PBE) \cite{pbe-1996} method to compare the band gap of an unstrained monolayer  MoS$_2$ with the work of Woo \textit{et al.} \cite{Woo-2016}. Our results show that the band gap decreases at a much slower rate than reported in the work by  Trainer \textit{et al.} \cite{DJ_Trainer_MoS2}. 

\begin{figure}
    \centering
    \includegraphics[width=\columnwidth]{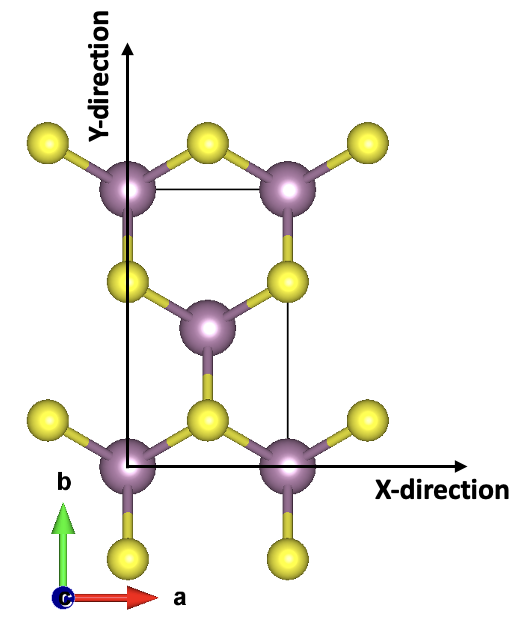}
    \caption{Orthorhombic unit cell of monolayer MoS$_2$. Only the S (yellow) atoms in the plane above the plane of the Mo (purple) atoms can be seen. Those in the plane below are obscured by those above. The lattice constant \textbf{a} is the length of the orthorhombic unit cell in the x-direction and \textbf{b} in the y-direction.}
    \label{fig:rect_cell}
\end{figure}
\maketitle
\section{Computational Methods} 
We used the orthorhombic unit cell as shown in  Figure \ref{fig:rect_cell} for our calculations. The orthorhombic unit cell helped to strictly maintain a Poisson's ratio of 0.44. The orthorhombic unit cell comprises two Mo atoms and four S atoms. The HSE06 and PBE calculations were performed using the Vienna Ab initio Simulation Package  (\textsc{vasp}) \cite{vasp-kresse-1994,vasp-kresse-1999}. For HSE06 and PBE calculations, we used a 16 x 16 x 2 $\Gamma$-centered k-grid to sample the Brillouin zone. We used a 400 eV cutoff energy for the plane-wave basis for HSE06 and 520 eV for PBE. Initially, we relaxed the lattice parameters \textbf{a} and \textbf{b} and the internal co-ordinates of ions while keeping \textbf{c} fixed at 10 Å to make a three-dimensional super cell. To apply strain, we utilized the relaxed structure and adjusted the lattice parameters  \textbf{a} and \textbf{b} to achieve the desired strain, followed by the relaxations of internal co-ordinates of ions. During the strain application, for HSE06 and G0W0@PBE calculations, we maintained the Poisson's ratio of 0.44 to remain consistent with experiment \cite{DJ_Trainer_MoS2}. If $\epsilon_x$ represents the strain along the x-direction, then the strain along the y-direction ($\epsilon_y$) would be $-0.44\epsilon_x$. If \textbf{a} and \textbf{b} are the lattice vectors before applying strain, and \textbf{a}$_{new}$ and \textbf{b}$_{new}$ are new lattice vectors after applying strain, then:
\begin{equation}
\mathbf{a}_{\text{new}}=\mathbf{a}\left(1+\frac{\epsilon_x}{100}\right)
\end{equation}
\begin{equation}
\mathbf{b}_{\text{new}}=\mathbf{b}\left(1+\frac{\epsilon_y}{100}\right).
\end{equation}

The G0W0 \cite{Deslippe2012,Hybertsen1986} calculations were conducted in BERKELEYGW \cite{Deslippe2012} by pairing with QUANTUM ESPRESSO (QE) \cite{Giannozzi1986}. The wavefunction energy cutoff is 70 Ry ($\sim$950 eV). The energy cutoff for the dielectric matrix ($\epsilon$) is 18 Ry ($\sim$240 eV). A k-point mesh of 20 × 12 × 1 is used for the orthorhombic cell. The band number for summation is 220. The correction of the exact static remainder and the 2D slab Coulomb truncation were used. The pseudopotentials used in QE are ONCVPSP (Optimized 
Norm-Conserving Vanderbilt Pseudopotential) fully-relativistic version. For atom Mo, the valence electrons are 4s$^2$/4p$^6$/4d$^5$/5s$^1$. For atom S, the valence electrons are 3s$^2$/3p$^4$.

The different k-meshes used in the different codes are both essentially converged for the band gap.  There would be no difference between one and two k-points in the direction out of the plane if the out-of-plane cell dimension were large enough. Increasing this cell dimension from 10 Å to 15 Å in VASP increased the PBE gap by only 0.05 eV.

\begin{table*}[ht]
  \centering
    \begin{tabular}{ccccccccccccc}
    \hline
    \hline
          &       & \multicolumn{3}{c}{HSE06} &       & \multicolumn{3}{c}{G0W0@PBE}+SOC &       & \multicolumn{3}{c}{Experiment} \\\cline{2-5} \cline{7-9} \cline{11-13}
    strain ($\epsilon_x$)\% &       & Band gap  &       & $\Delta$ &       & Band Gap &       & $\Delta$ &       & Band Gap &       & $\Delta$ \\
    \hline 
    0.00  &       & 2.30  &       & 0.00  &       & 2.64  &       & 0.00  &       & 2.34  &       & 0.00 \\
    1.00  &       & 2.25  &       & -0.05 &       & 2.60  &       & -0.04 &       & 1.94  &       & -0.40 \\
    2.50  &       & 2.15  &       & -0.15 &       & 2.53  &       & -0.11 &       & 1.34  &       & -1.00 \\
    4.00  &       & 2.01  &       & -0.29 &       & 2.46  &       & -0.19 &       & 0.74  &       & -1.60 \\
    \hline
    \hline
    \end{tabular}%
      \caption{The computed and experimental band gaps (in eV) and changes ($\Delta$) in the band gap relative to the gap at zero strain of monolayer MoS$_2$.  The HSE06 and G0W0@PBE results correspond to the orthorhombic unit cell shown in Fig. 1. Experimental values from Ref. \cite{DJ_Trainer_MoS2}.}
  \label{tab:bg_theo_rectcell}%
\end{table*}%
\begin{table*}[ht]
  \centering
    \begin{tabular}{cccccccccc}
    \hline
    \hline
    strain ($\epsilon_x$) \% &       &       & Band gap &       &       & $\Delta$ &       &       & Lattice constant \\
    \hline
    0.00  &       &       & 1.65  &       &       & 0.00  &       &       & 3.18 \\
    1.00  &       &       & 1.45  &       &       & -0.20 &       &       & 3.21 \\
    2.50  &       &       & 1.15  &       &       & -0.50 &       &       & 3.26 \\
    4.00  &       &       & 0.82  &       &       & -0.83 &       &       & 3.31 \\
    6.00  &       &       & 0.45  &       &       & -1.20 &       &       & 3.37 \\
    8.00  &       &       & 0.17  &       &       & -1.48 &       &       & 3.43 \\
    10.00 &       &       & 0.02  &       &       & -1.63 &       &       & 3.50 \\
    \hline
    \hline
    \end{tabular}%
      \caption{band gap vs. strain and lattice parameter (in \AA ) calculated using PBE for the orthorhombic cell shown in Fig. \ref{fig:rect_cell}.  Here, we applied uniform biaxial tensile strain.}
  \label{tab:PBE_bg_vs_strain}%
\end{table*}%

\maketitle
\section{Results and Discussion}
First, we calculated the band gap of the unstrained monolayer MoS$_2$. For this, we relaxed lattice parameters \textbf{a} and \textbf{b} and internal co-ordinates of the ions. Then, we used the relaxed structure to obtain band gaps. The lattice parameter \textbf{a} for the unstrained relaxed structure is 3.15Å for HSE06 and 3.18 Å for PBE. Our lattice parameter agrees with references \cite{Johari-2012, Pela2024}. We obtained band gaps of 1.65 eV, 2.30 eV, and 2.64 eV using PBE \cite{pbe-1996}, HSE06, and G0W0@PBE methods, respectively, for unstrained monolayer MoS$_2$. These results agree with Refs. \cite{Johari-2012, Pela2024} for the unstrained monolayer MoS$_2$. Our HSE06 band gap agrees with the experimental band gap of 2.34 eV obtained using STM \cite{DJ_Trainer_MoS2}, however the G0W0@PBE band gap is slightly larger. Trainer \textit{et al.} \cite{DJ_Trainer_MoS2} proposed the relation between the band gap ($E_\mathrm{g}$) and strain ($\epsilon_x$) as in equation \ref{eq:eg_vs_strain_exp}. Table \ref{tab:bg_theo_rectcell} summarizes the band gap and change in band gap vs. strain based on relation \ref{eq:eg_vs_strain_exp} in the last two columns. 
\begin{equation}
E_{\text{g}}=\left(-0.401\epsilon_x+2.343\right)\ eV.
\label{eq:eg_vs_strain_exp}
\end{equation}
\begin{figure}
    \centering
    \resizebox{0.5\textwidth}{!}{\includegraphics{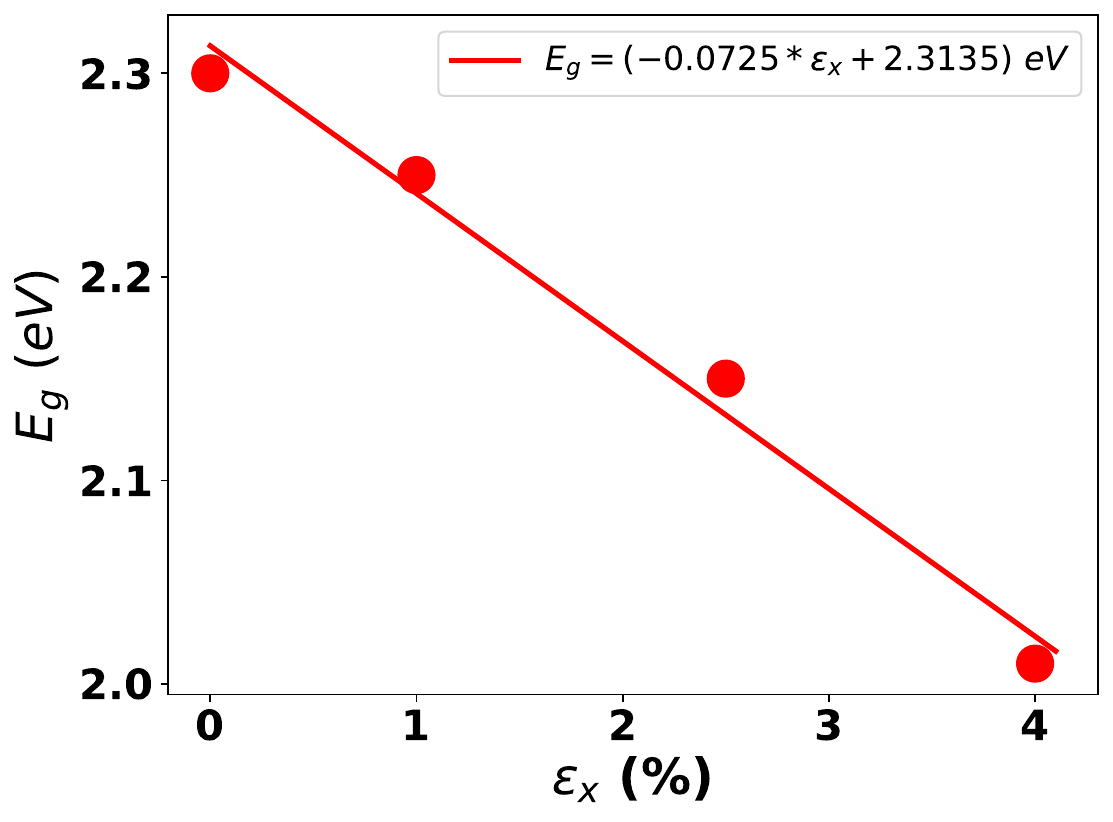}}
    \caption{Change in the band gap $E_\mathrm{g}$ as a function of strain $\epsilon_x$ along the x-direction using the orthorhombic unit cell of monolayer MoS$_2$ based on the HSE06 functional. We applied strain such that Poisson's ratio remains 0.44.}
    \label{fig:Eg_vs_ex}
\end{figure}
\begin{table*}[htbp]
  \centering
    \begin{tabular}{ccccccccccccc}
    \hline
    \hline
          &       & \multicolumn{4}{c}{Poisson's ratio: 0.25} &       &       &       & \multicolumn{4}{c}{Poisson's ratio: 0.44} \\\cline{2-7} \cline{9-13} 
    strain ($\epsilon_x$) \% &       & Band gap &       &       & $\Delta$ &       &       &       & Band gap &       &       & $\Delta$ \\
    \hline
    0.00     &      & 1.65   &       &       &  0.00     &       &       &       & 1.65   &       &       & 0.00 \\
    1.00     &       & 1.60  &       &       & -0.05    &       &       &       & 1.60   &       &       & -0.05 \\
    2.50     &       & 1.45  &       &       & -0.20    &       &       &       & 1.50   &       &       & -0.15 \\
    4.00     &       & 1.30  &       &       & -0.35   &       &       &       & 1.40   &       &       & -0.25 \\
    \hline
    \hline
    \end{tabular}%
      \caption{PBE band gap (in eV) vs. strain evaluated for Poisson's ratios 0.25 and 0.44. }
  \label{tab:PBE_bg_change}%
\end{table*}%
\begin{figure*}
    \centering
    \resizebox{1.0\textwidth}{!}{\includegraphics{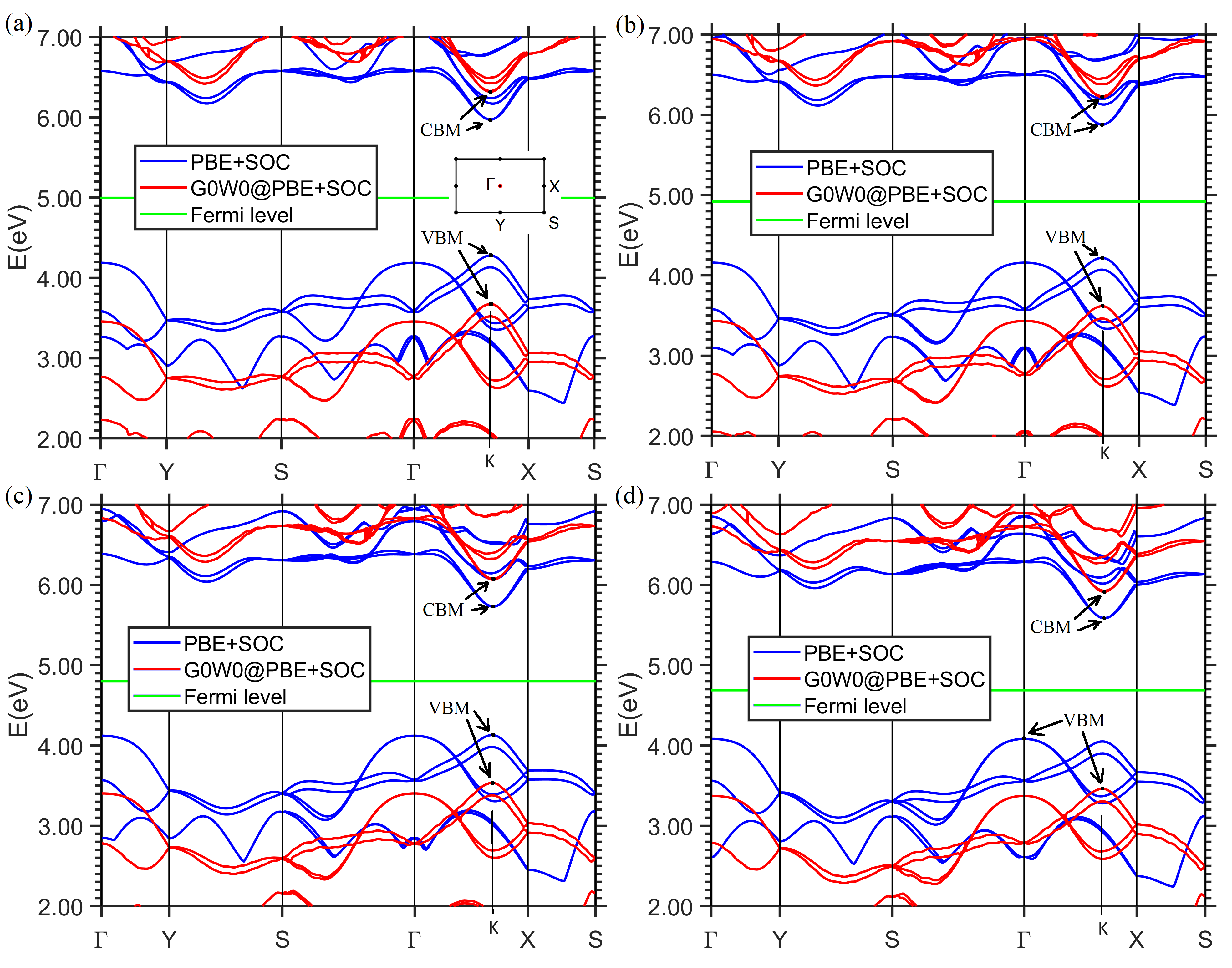}}
    \caption{Band structure of (\textbf{a}) unstrained, (\textbf{b}) 1\%, (\textbf{c}) 2.5 \%, and (\textbf{d}) 4\% strained monolayer MoS$_2$ calculated using a orthorhombic cell with PBE+SOC and G0W0@PBE+SOC methods. We applied strain such that Poisson's ratio remains 0.44. The plot shows unstrained monolayer MoS$_2$ to be a direct band gap material, and 4\% strained monolayer MoS$_2$ to be an indirect band gap material at the PBE level. However, it is still almost direct at the G0W0@PBE level. (By K we indicate the point along the line from $\Gamma$ to X  that is closest to the symmetry point K of the hexagonal zone.)}
    \label{fig:bs_rect_cell}
\end{figure*}
To compare with the experimental results, we calculated the band gap of monolayer MoS$_2$ theoretically using the HSE06 and G0W0@PBE methods using the orthorhombic unit cell as shown in Fig. \ref{fig:rect_cell}. We used  an orthorhombic unit cell to easily maintain the Poisson's ratio of 0.44 determined by Trainer \textit{et al.} \cite{DJ_Trainer_MoS2}. We applied $\epsilon_x$ $\%$ strain along the x-direction and $-0.44 \epsilon_x \%$ strain along the y-direction.  As depicted in Figure \ref{fig:rect_cell}, we aligned the x-direction along lattice vector \textbf{a} and the y-direction along lattice vector \textbf{b} of the orthorhombic cell. Table \ref{tab:bg_theo_rectcell} summarizes the band gap and the change in the band gap evaluated using the HSE06 and G0W0@PBE methods. The change in the band gap at any strain $\epsilon_x$ is evaluated relative to the band gap of the unstrained sample. Additionally, we exchanged the x and y directions (i.e., the x-direction aligned with the lattice vector \textbf{b} and y-direction along lattice vector \textbf{a} ) and computed the band gap and the change in the band gap versus strain using the HSE06 method. However, we did not observe any difference in the results. To compare our result with the result of Ref. \cite{DJ_Trainer_MoS2}, we applied up to  4\% strain.  Table \ref{tab:bg_theo_rectcell} clearly shows that the change in the band gap vs. strain predicted by HSE06 and G0W0@PBE is much smaller than in the experiment \cite{DJ_Trainer_MoS2}. The change in the band gap predicted by G0W0@PBE for 2.5\% strain is 0.11 eV. This agrees with the result of Ref.  \cite{Conley2013}, but not with the result of Ref. \cite{DJ_Trainer_MoS2}.  We also quantified the change in the band gap per \% of strain. For this, we applied a linear fit to HSE06 band gap vs. strain data as shown in Figure \ref{fig:Eg_vs_ex}. The fit shows that the band gap decreases at a rate of 73 meV/\% strain, which is much smaller than in the experimental rate of 400 meV/\% strain \cite{DJ_Trainer_MoS2}. A similar fit for G0W0@PBE shows a decrease of band gap at the rate of 45 meV/\% strain.

The discrepancy between theory and experiment \cite{DJ_Trainer_MoS2} in the band gap change vs. strain in monolayer MoS$_2$ is significant and needs to be resolved. Ref. \cite{DJ_Trainer_MoS2} mentions that the strain is not uniformly distributed in the sample, which is also evident from Figure 2e of Ref. \cite{DJ_Trainer_MoS2}. Ref. \cite{DJ_Trainer_MoS2} also reports the slipping of the sample with respect to the substrate. It appears that the strain is not uniformly transferred from the substrate to the film. If the strain transfer process induced some ripple effect in the sample, then it is possible that the sample has regions with unusually high strain.

A more interesting possibility is that the graphite surface, strained by the monolayer, has a strong proximity effect on the bandgap of the monolayer. While this could probably be tested by calculations, it is beyond the scope of this work.

We first applied the theoretical strain by considering that monolayer MoS$_2$ has a Poisson's ratio of 0.44. However, the PBE calculations of Refs. \cite{Kang2015,Xiong2015} show much smaller values of Poisson's ratio of 0.25 and 0.267, respectively, for free-standing monolayer MoS$_2$. Therefore, we also used the Poisson's ratio values of 0.25 and 0.44 from the PBE Ref. \cite{Kang2015} and from the experimental Ref. \cite{DJ_Trainer_MoS2}, respectively, and calculated the effect of biaxial strains maintaining these Poisson's ratios using the PBE functional. Table \ref{tab:PBE_bg_change} shows the results. As expected, the band gap decreases faster for the Poisson's ratio of 0.25 compared to that of 0.44. We also quantified the rate of decrease of the band gap by applying a linear fit to the band gap vs. strain data of Table \ref{tab:PBE_bg_change}.
The fitted equations for Poisson's ratio 0.25 and 0.44 are:
\begin{equation}
    E_\mathrm{g}=-0.0898\epsilon_x\ + \ 1.6684\ eV
\end{equation}
and
\begin{equation}
    E_\mathrm{g}=-0.0633\epsilon_x\ + \ 1.6561\ eV,
\end{equation}
respectively.
The rate of decrease of the band gap for the Poisson's ratio of 0.44 is 63 meV/\% strain, which is not very different from the HSE06 result of 73 meV/\% strain, and for the Poisson's ratio of 0.25 is 90 meV/\% strain. This also indicates that the value used for Poisson's ratio affects the rate of change of the band gap, but not dramatically. Our calculated values are not surprisingly different from the values of 59 and 94 meV/\% strain for direct and indirect fundamental band gap, respectively, calculated using the GW0-BSE method in Ref. \cite{Conley2013}. Ref. \cite{Conley2013}  also reports the rate of decrease of the optical band gap using photoluminescence spectroscopy to be nearly 45 meV/\% strain. Possible errors in Poisson's ratio do not seem to be the reason for the large discrepancy in band gap change with strain between theoretical and experimental results.
\begin{figure}
    \centering
    \resizebox{0.5\textwidth}{!}{\includegraphics{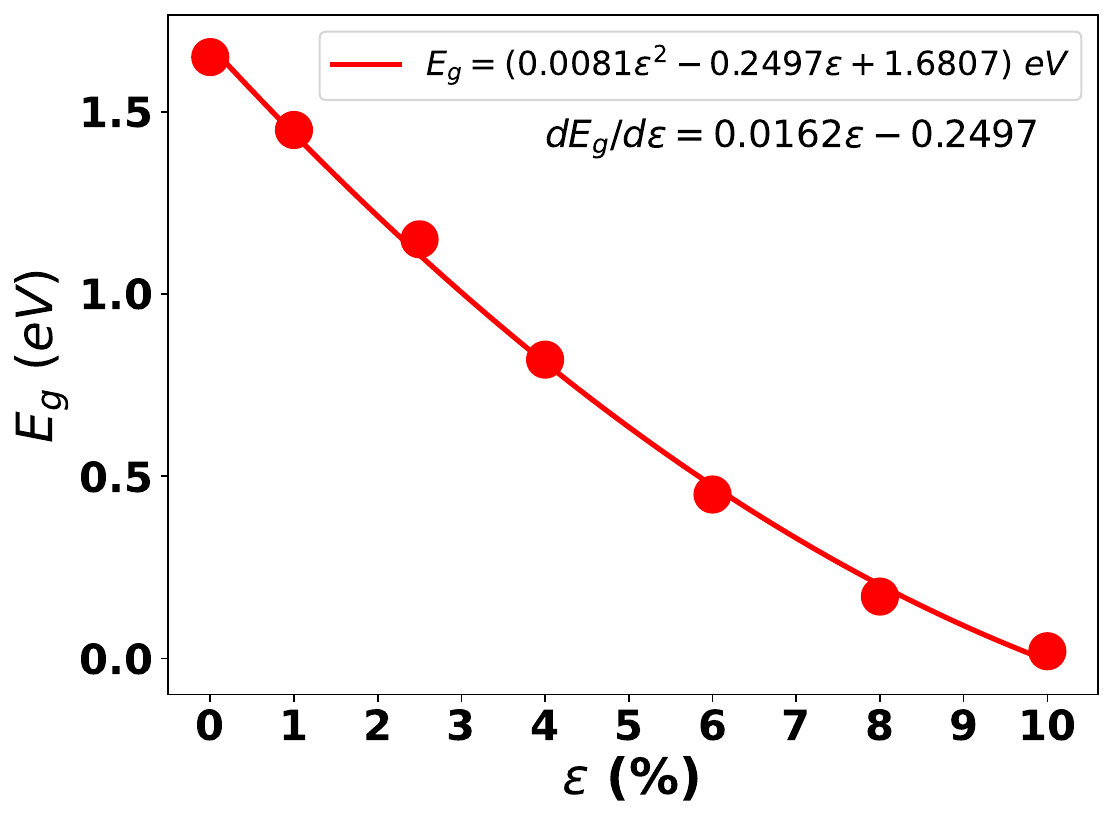}}
    \caption{PBE band gap $E_\mathrm{g}$ as a function of equal strain $\epsilon$ applied along both the x and y directions using a orthorhombic unit cell of monolayer MoS$_2$ (equivalent to a Poisson's ratio of -1).}
    \label{fig:PBE_bg_vs_strain}
\end{figure}

The band gap of 1H $\mathrm{MoS_2}$ monolayer decreases with increasing in-plane tensile strain, a trend opposite to that of some other transition metal dichalcogenides (TMDs), such as the 1T $\mathrm{TiSe_2}$ monolayer \cite{Diego2018}. The decreasing trend in $\mathrm{MoS_2}$ can be attributed to the effect on the band centers of the change in ligand field induced by the in-plane strain, the change of the widths of the $d$-orbital bands, and the change of $d$-orbital SOC splitting. A more detailed analysis is presented in the Supplementary Information (SI) \cite{SM2024}, which includes Refs.
\cite{2D-MoS2,Tang2023,Yoffe1969}.

To investigate the effect of spin-orbit 
correction (SOC), we also calculated the PBE+SOC and G0W0@PBE+SOC band structures of monolayer MoS$_2$ for no strain, 1.0\% strain, 2.5\% strain, and 4\% strain along the x-direction with a strain along the y-direction to maintain Poisson's ratio of 0.44. The SOC effect is not always significant for band structures of 2D materials. For example, the SOC effect on the bands around the Fermi level of monolayer $\mathrm{Ta_2Ni_3Te_5}$ was shown to be very small \cite{Guo2022}. For 1H WSe$_2$ monolayer, the SOC effect induces a strong spin-splitting ($\sim$ 0.5 eV) around the valence band maximum \cite{Retana2016}. Here, for the 1H MoS$_2$ monolayer, the SOC-induced band splitting near the valence band maximum is about 0.15 eV, consistent with a previous study \cite{Diana2013}. As shown in Figure \ref{fig:bs_rect_cell}a and \ref{fig:bs_rect_cell}b, both PBE+SOC and G0W0 show the unstrained  and 1.0\% strained monolayer MoS$_2$ to be a direct band gap semiconductor. At 2.5\% strain, PBE+SOC still shows a direct gap, while G0W0 shows that the valence band maximum (VBM) and the conduction band minimum (CBM) located in the $\Gamma$X line and close to X are slightly misaligned, showing a slightly indirect gap. At 4\% strain, PBE+SOC shows that the local VBM at $\Gamma$ has become the global VBM, yielding an indirect gap. Therefore, with increasing strain this material tends to change from a direct band gap at an indirect band gap material. Such an effect has been previously observed in Refs. \cite{Zeng2012,Conley2013,Johari-2012}.

Additionally, theory predicts that 10\% biaxial strain might close the band gap of MoS$_2$ \cite{Johari-2012,Yue-2012,Shi-2013}. We applied 10\% strain along the x-direction and -4.4\% strain along the y-direction to maintain the Poisson's ratio of 0.44 and found that HSE06 gives a band gap of 1.6 eV, which is much greater than 0 eV. We also applied equal biaxial tensile strains of 10\% along both the x and y directions and observed that HSE06  significantly reduces the band gap to 0.30 eV, but does not close it completely. However, PBE almost closes the band gap of monolayer MoS$_2$ for 10\%  equal biaxial tensile strain applied along both x and y directions. This result is consistent with the work in Ref. \cite{Johari-2012}. Table \ref{tab:PBE_bg_vs_strain} shows strain vs. band gap and lattice parameter \textbf{a} of monolayer MoS$_2$ obtained using the PBE functional for up to 10\% of equal biaxial tensile strains. Table \ref{tab:PBE_bg_vs_strain} agrees with Table 1 of Ref. \cite{Johari-2012}. We also plotted the PBE band gap vs. strain up to 10\% of strain as shown in Fig. \ref{fig:PBE_bg_vs_strain}. The plot is again similar to the plot in Ref. \cite{Johari-2012}. The plot shows that band gap is not a linear function of strain. The rate of decrease of band gap ($dE_\mathrm{g}/d\epsilon$) is:
\begin{equation}
    dE_\mathrm{g}/d\epsilon=0.0162\epsilon-0.2497 \ eV/\%,
\end{equation}
where $\epsilon$ is the equal biaxial tensile strain applied. The band gap decrease becomes slower with increase in strain.
\maketitle
\section{Conclusion}
The ability to control the electronic properties of a material by applying strain is of critical importance, as it widens the application of 2D materials into flexible electronics and next-generation strain engineering devices. 2D MoS$_2$ occupies a significant role in these applications due to its important properties such as semiconducting nature, high tensile strength and flexibility, and high carrier mobility. To be used in the flexible electronics and next-generation strain engineering devices, it is crucial to understand how strain affects the electronic properties of this material. 

In this work, we theoretically studied the band gap and its change in free-standing monolayer MoS$_2$ using the HSE06 and G0W0 methods to compare with the experimental finding for monolayer MoS$_2$  on a mildly curved graphite substrate \cite{DJ_Trainer_MoS2}. We quantified the rate of decrease of band gap with strain (that maintains Poisson's ratio of 0.44) for monolayer MoS$_2$. We found the rate of decrease of band gap to be 63 meV/\% strain for PBE, 73 meV/\% strain for HSE06 and 45 meV/\% strain for G0W0@PBE, which are much smaller than the experimental result of 400 meV/\% strain \cite{DJ_Trainer_MoS2}. The discrepancy is important to resolve. It suggests the possibility of an unexpectedly strong effect of the substrate on the strain dependence of the gap. Furthermore, we demonstrated that PBE and G0W0@PBE predict monolayer MoS$_2$ to be a direct band gap material, with the potential to transition into an indirect band gap material under strain. We also quantified the rate of decrease of band gap with strain (that maintains Poisson's ratio of 0.25) to be 90 mev/\% strain at PBE level.  Additionally, we observed that 10\% equal biaxial tensile strain can transform  MoS$_2$ from a semiconductor to a metal at the PBE level, but not at the HSE06 level.

Finally, we briefly comment on the reason why the HSE06 energy gap in the bandstructure of a solid is typically larger and more realistic than the PBE gap. The PBE density functional for the exchange-correlation energy is implemented in the original Kohn-Sham \cite{kohn1965self} scheme in which the exchange-correlation potential is a function of position that multiplies an orbital, while the HSE06 hybrid of PBE with a fraction of exact exchange depends explicitly on the occupied orbitals and is implemented in a generalized \cite{Seidl1996} Kohn-Sham scheme in which the exchange-correlation potential is a non-local operator. PBE reasonably predicts the bandstructure gap that would be predicted by a position-dependent multiplicative exchange-correlation potential that yields the exact electron density in a solid \cite{Levy1983,Sham1983}, while HSE06, to the extent that it correctly predicts ionization energy minus electron affinity for the solid from total energy differences, also predicts the correct fundamental gap from its bandstructure \cite{Perdew2017}. The difference between the HSE06 and PBE gaps is an estimate of the discontinuity \cite{Levy1983,Sham1983,Perdew2017,Perdew1982} of the exact Kohn-Sham potential in an insulating solid as the electron number crosses the number needed to fill a band. Another way to understand this is to note that the electron self-energy that yields the exact quasi-particle spectrum must be a nonlocal operator \cite{Sham1983}, like HSE06 and unlike the PBE  and  the exact Kohn-Sham exchange-correlation potentials.
\maketitle
\section{Acknowledgements}
The work of RKS was supported by the Department of Energy, Office of Science, Basic Energy Sciences, Materials Chemistry Materials Science and Engineering, under grant DE-DC0023356. The work of RKS and HT was partially supported by the Temple Center for Research on Energy Materials through the Catalytic Collaborative Funding Initiative sponsored by the Office of the Vice President for Research at Temple. The work of JPP was supported by DOE grant DE-SC0023356, and by the National Science Foundation, Division of Materials Research, Condensed Matter and Materials Theory, under grant DMR-2344734. We thank NERSC for computational
resources. We thank Professor Maria Iavarone for helpful discussions and suggestions.
%\maketitle
%\section{Supplementary Information}
%\textcolor{red}{The Supplementary Information includes a possible explanation for the decrease in band gap with tensile strain, and a link to the inputs for the VASP, Quantum Espresso, and Berkeley GW codes.}


\clearpage % Start a new page
\onecolumngrid
\noindent
\textbf{Supplementary Information}\\\\
\noindent
\justifying
\textbf{Possible explanation for the decrease in band gap with strain}\\

\begin{figure}
\renewcommand\thefigure{S1}
    \centering
    \resizebox{0.4\textwidth}{!}
    {\includegraphics{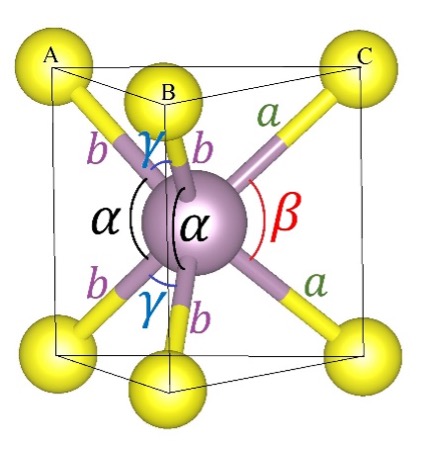}}
    \caption{The structure of the trigonal prism of the 1H MoS$_2$ monolayer. The direction AB is that of the x axis, and the y axis is perpendicular to AB, as in Figure 1. The three angles $\alpha$, $\beta$, and $\gamma$, and bond lengths a and b are denoted in different colors for clarity. 
    }
    \label{fig:vvv}
\end{figure}

In the 1H MoS$_2$ monolayer, one Mo atom and its six adjacent S atoms form a trigonal prism (see Figure \ref{fig:vvv}), and in the ligand field theory, the d-orbitals of the transition metal atoms split into three groups of non-bonding levels [5]  %\cite{Chhowalla2013}. 
The lower split group is mainly of $d_{z^2}$ character and is filled, and the fundamental gap is formed by this filled group and the unoccupied middle group.

To understand the trend in the change of gap with applied tensile strain, we analyze the band widths of the relevant d-orbitals and the spin-split of the valence band around the K point, and the results are listed in Table \ref{tab:TableX}. As can be seen, the spin-orbit coupling (SOC) induced spin-split is almost unchanged with an increase in strains, implying that the SOC effect may not be obviously influenced by the strains. This is unlike the case in the bent WSe$_2$ nanoribbons, where the non-uniform bending strain induced flexoelectric effect can dramatically increase the SOC effect [30] % \cite{Tang2023}. 

The band widths of the d-orbital derived valence bands just below the Fermi level show a roughly unchanged trend with strains, and those of conduction bands just above the Fermi level show a slightly increasing trend with strains, as seen in Table \ref{tab:TableX}. The relevant d-orbital band broadening is believed to be the main reason for the gap's decreasing trend from 2H-MoS$_2$, to 2H-MoSe$_2$, and to 2H-MoTe$_2$, with increasing atomic number of the chalcogen [31] %\cite{Yoffe1969}. 
However, the total band width broadening of bands around the Fermi level cannot completely explain the band gap decrease of monolayer MoS$_2$ with in-plane strains here. Comparing the case of strain $\epsilon_x=4\%$ with the non-strained one, G0W0 gives a gap decrease of 190 meV, see Table 1 in the main text, while the change of the total band broadening is only about 98 meV, see Table \ref{tab:TableX}. So, the band gap decrease of monolayer MoS$_2$ with strains should also involve the relevant band-center shifts. 

The gap is also related to the separation between the band-centers of the above-mentioned lower filled split group and the middle unoccupied group. The separation is determined by the ligand field symmetry and strength. Figure \ref{fig:vvv} shows the trigonal prism of 1H MoS$_2$, with the relevant angles and bond lengths denoted. The AB direction is in the x axis, where the strain $\epsilon_x$ is applied, and the y axis is perpendicular to AB. The three angles $\alpha$, $\beta$, and $\gamma$, and bond lengths a and b in the prism under different strains are listed in Table \ref{tab:TableY}. For the unstrained case, the angles $\alpha$ and $\beta$ are equal, and the bond lengths a and b are also equal. The prism has a three-fold rotational symmetry. This symmetrical configuration may result in the largest ligand field splitting between the lower filled and the middle unoccupied groups. With increasing strains, the angle $\alpha$ decreases, and the angle $\gamma$ increases, while angle $\beta$ remains almost unchanged, and the bond length a decreases, but b increases, due to the stretching in the x axis and compression in the y axis. The prism loses the three-fold rotational symmetry, and the ligand field also changes its symmetry and strength, resulting in a decrease in the splitting of the lower and the middle groups, hence a decrease in the gap.

\begin{table*}[ht]
\renewcommand\thetable{S1}
  \centering
    \begin{tabular}{ccp{1.5cm}p{1.5cm}p{1.5cm}p{1cm}c|cp{1.5cm}p{1.5cm}p{1.5cm}p{1cm}}
    \hline
    \hline
          &       & \multicolumn{4}{c}{G0W0@PBE+SOC} &     &  & \multicolumn{4}{c}{PBE+SOC} \\\cline{3-6} \cline{9-12}
     &       & Band width\_v& Band width\_c & TBWB & Spin split & & & Band width\_v    & Band width\_c  &  TBWB   &  Spin split \\
    \hline 
    no strain  &  & 1.205   & 0.731  &  0.000  &   0.152 & & & 1.061  & 0.648    & 0.000   & 0.146 \\
    $\epsilon_x$=1.0\% & & 1.205   & 0.751  &  0.020  &   0.154 & & & 1.066  & 0.667    & 0.025   & 0.147 \\
    $\epsilon_x$=2.5\% & & 1.203   & 0.794  &  0.060  &   0.155 & & & 1.057  & 0.707    & 0.056  & 0.148 \\
    $\epsilon_x$=4.0\% & & 1.189   & 0.846  &  0.098  &   0.156 & & & 1.067  & 0.754   & 0.112  & 0.149 \\
    \hline
    \hline
    \end{tabular}%
      \caption{Contributions to the decrease of bandgap in monolayer $\mathrm{MoS_2}$ with increasing in-plane tensile strain. TBWB stands for total band width broadening. The band widths and broadening of the d-orbital derived bands around the Fermi level and the spin-splits for 1H $\mathrm{MoS_2}$ monolayer under different strain conditions. In strained cases, the strain along the y axis is $\epsilon_y=-0.44 \epsilon_x$. Band width$_v$ is the band width of the four valence bands just below the Fermi level. Band width$_c$ is the band width of the two conduction bands just above the Fermi level. Total band width broadening is referenced to the non-strained case. The spin-split is the one at the top of the valence band around the K point. Unit of entries: eV.}
  \label{tab:TableX}%
\end{table*}%
\begin{table*}[ht]
\renewcommand\thetable{S2}
  \centering
    \begin{tabular}{ccccccccccc}
    \hline
    \hline
        &   &   $\alpha$    &   & $\beta$& & $\gamma$ & & a & & b\\
    \hline
    no strain  &  & 80.850 & & 80.850  & & 82.488 & & 2.386 & & 2.386 \\
    $\epsilon_x$=1.0\% &  & 80.483 & & 80.875  & & 83.251 & & 2.382 & & 2.392 \\
    $\epsilon_x$=2.5\% &  & 79.820 & & 80.876  & & 84.427 & & 2.374 & & 2.400 \\
    $\epsilon_x$=4.0\% &  & 79.193 & & 80.867  & & 85.593 & & 2.367 & & 2.408 \\
    \hline
    \hline
    \end{tabular}%
      \caption{The three angles $\alpha$, $\beta$,  and $\gamma$, and bond lengths a and b in the trigonal prism, depicted in Figure \ref{fig:vvv}, of 1H MoS$_2$ monolayer under different strain conditions. In strained cases, the strain along the y axis is $\epsilon_y=-0.44 \epsilon_x$. Units: angle: degree; bong length: angstrom.}
  \label{tab:TableY}%
\end{table*}%

\noindent
\textbf{GitHub repository for Input Files}

You can download the input files from this link:
\href{https://github.com/Rajkishor46/MoS2Inputs}{https://github.com/Rajkishor46/MoS2Inputs}.
\end{document}